\documentclass[journal=apchd5,manuscript=letter]{achemso}

\usepackage[version=3]{mhchem} 
\usepackage[T1]{fontenc}       
\usepackage{SIunits}

\author{Andreas W. Schell}
\affiliation[Kyoto University]
{Department of Electronic Science and Engineering, Kyoto University, 615-8510 Kyoto, Japan}
\email{aws.kyoto@gmail.com}
\author{Hideaki Takashima}
\affiliation[Kyoto University]
{Department of Electronic Science and Engineering, Kyoto University, 615-8510 Kyoto, Japan}
\author{Toan Trong Tran}
\affiliation[UTS]
{School of Mathematical and Physical Sciences, University of Technology Sydney, Ultimo, New South Wales 2007, Australia}
\author{Igor Aharonovich}
\affiliation[UTS]
{School of Mathematical and Physical Sciences, University of Technology Sydney, Ultimo, New South Wales 2007, Australia}
\author{Shigeki Takeuchi}
\affiliation[Kyoto University]
{Department of Electronic Science and Engineering, Kyoto University, 615-8510 Kyoto, Japan}
\email{takeuchi@kuee.kyoto-u.ac.jp }

\title{Coupling quantum emitters in 2D materials with tapered fibers}


\begin{document}

\begin{tocentry}
\includegraphics{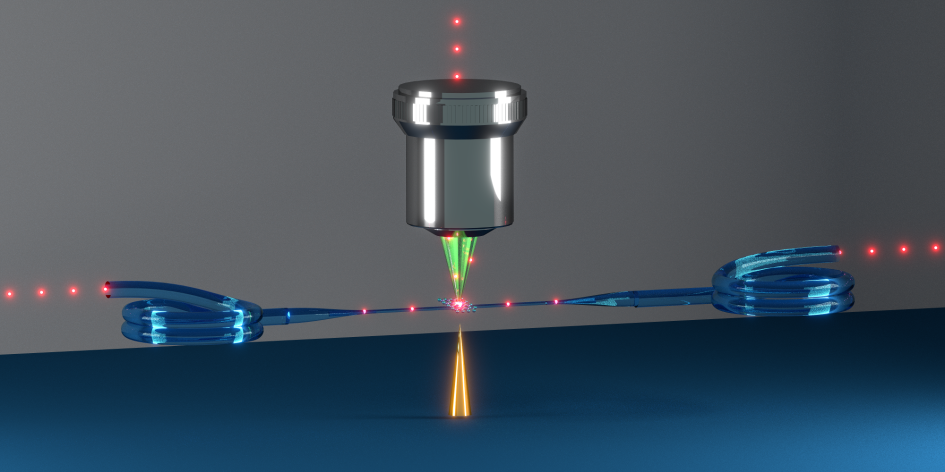}
\end{tocentry}

\begin{abstract}
Realization of integrated photonic circuits on a single chip requires controlled manipulation and integration 
of solid-state quantum emitters with nanophotonic components. 
Previous works focused on emitters embedded in a three-dimensional crystals -- such as nanodiamonds or quantum dots.
In contrast, in this work we demonstrate coupling of a single emitter in a two-dimensional (2D) material, 
namely hexagonal boron nitride (hBN), with a tapered optical fiber and find a 
collection efficiency of the system is found to be 10~\%.
Furthermore, due to the single dipole character of the emitter, we were able to analyse the
angular emission pattern of the coupled system via back focal plane imaging.   
The good coupling efficiency to the tapered fiber even allows excitation and 
detection in a fully fiber coupled way yielding a true integrated system.
 Our results provide evidence of the feasibility 
to efficiently integrate quantum emitters in 2D materials with photonic structures. 

\end{abstract}

Atomically thin two-dimensional (2D) materials are garnering major attention in a variety of emerging applications 
in nanophotonics and optoelectronics~\cite{Mak2016,Novoselov2016} spanning low threshold lasing~\cite{Wu2015}, 
sub-diffraction imaging~\cite{Caldwell2014,Dai2014}, valleytronics~\cite{Yang2016,Sie2015}, and efficient light 
emitting diodes~\cite{Clark2016,Withers2015}. 
In particular, hexagonal boron nitride (hBN) has recently emerged as a promising layered material that 
hosts optically stable single photon emitters that operate at room temperature~\cite{Tran2016,Bourrellier2016,Jungwirth2016,Chejanovsky2016}. 
These defects have a narrow emission linewidth from the ultraviolet to the near infrared spectral 
range, are bright, and can be excited by two-photon excitation~\cite{Schell2016}. These properties make  
them promising candidates for applications in (bio)-sensing~\cite{Helmchen2005}, 
nanophotonics~\cite{Gaponenko2010}, and quantum information science~\cite{Weber2010}.

However, in order to employ these emitters in applications, e.g., as single photon sources, 
it is necessary to efficiently extract the emitted photons~\cite{Aharonovich2016}.
While usage of solid immersion lenses~\cite{Mansfield1990} and 
dielectric antennas~\cite{Lee2011} has been demonstrated, these approaches do not offer a path to 
scalability and suffer from complicated output modes. 
A promising direction to overcome this problem is using near-field coupling to 
tapered optical fibers~\cite{Yalla2012,Fujiwara2011,Liebermeister2014}, where 
the photons get directly emitted into a guided mode inside the fiber. 
In these geometries, coupling efficiencies of over 20~\% can be reached and even higher efficiencies are  
possible by combining the tapered fibers with optical cavities, such as nanofiber Bragg cavities 
(NFBCs)~\cite{Yalla2014,Schell2015,Takashima2016}.

In this letter, we demonstrate unprecedented results of coupling quantum emitters embedded in layered 
hBN to a 
tapered optical fiber. 
We show efficient collection of light emitted by a single photon emitter 
into the optical fiber and prove that the quantum nature of light is maintained throughout 
when the light is guided through the fiber. We further analyze the collection efficiency in detail 
by taking into account the non-isotropic emission of a dipole emitter.

\begin{figure}
	\centering
		\includegraphics{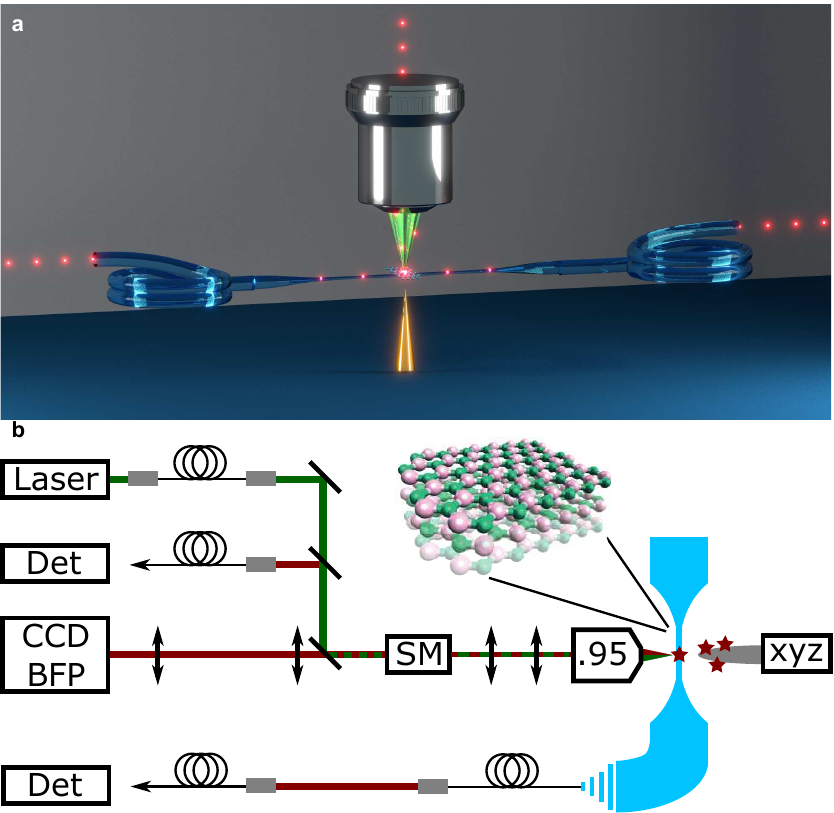}
		\caption{\textbf{Setup used in the experiment.}
		\textbf{a}, Illustration of the experiment. 
		\textbf{b},
		The setup used in this experiment consists of a home built confocal microscope 
		equipped with a piezo-driven three-axis manipulator (xyz) for attaching hBN flakes to 
		a tapered optical fiber (colored blue). The microscope uses an objective lens with 
		a numerical aperture of 0.95 and raster scanning is achieved using a scanning mirror assembly (SM).
		Different lasers (pulsed or cw) can be fiber coupled to the microscope and detection either occurs via 
		a CCD camera for imaging the sample or the back focal plane (BFP) of the objective. Detection (Det)
		of the light is done with interchangeable  modules for registering intensity, spectrum, and 
		intensity correlations. Fiber coupling of the modules makes their use with the microscope as 
		well as with the fiber possible.
		Light inside the fiber is coupled out for filtering and then coupled to another fiber and sent to the 
		detection units (Det).
		hBN flakes are depicted in the rendering (rose is boron, green is nitrogen) and as stars.
		}
		\label{fig:setup}
\end{figure}

A sketch of the setup is shown in Figure~\ref{fig:setup} (see Methods for detailed description).
The setup consists of a home-built beam-scanning confocal microscope equipped with a spectrometer and 
avalanche photodiodes for photon correlation measurements. 
The emitters in hBN (see Methods) are excited using pulsed or continuous wave lasers at a wavelength of \unit{532}{\nano\meter}.
A tapered optical fiber with a radius of \unit{320}{\nano\meter} is positioned in front of the objective lens with a 
numerical aperture (NA) of 0.95. One end of the fiber is spliced to a longer fiber from 
which the light is coupled out in order to perform filtering before it is sent to the detectors.

To transfer the hBN flakes that host the quantum emitters onto the tapered fiber 
a sharp tungsten tip driven by a piezo positioner is used~\cite{Fujiwara2016}. The hBN flakes  
are lifted off the silicon substrate with the tip. Subsequently, the 
flakes are then placed onto the fiber by approaching the tungsten tip to the 
tapered region of the fiber. After successful transfer of an hBN flake, 
room-temperature optical measurements were carried out. 

\begin{figure}
	\centering
		\includegraphics{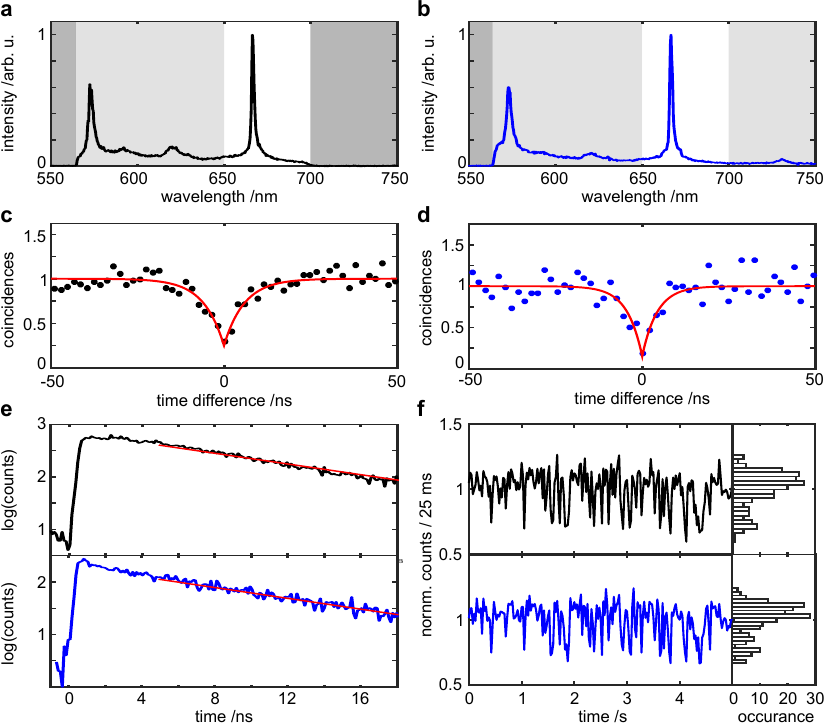}
		\caption{\textbf{Characterization of a deposited hBN flake.}
		\textbf{a,b}, 
    spectrum of the light collected through the microscope objective and the fiber, respectively.
		Dark gray wavelength ranges are blocked by filters. In the measurements in 
		panel f the light gray ranges are also blocked. For measurements in the other panels,
		a \unit{10}{\nano\meter} bandpass filter was used to suppress all light but the 
		light coming from the peak at \unit{666}{\nano\meter}.
		\textbf{c,d}, antibunching measurements of the light collected through the 
		objective and through one end of the fiber, respectively. The pronounced dips 
		at zero time delay prove the single photon nature of the light. Values at zero time 
		difference as extracted from the fits (red lines) are $g^{(2)}(0)=0.24$ and 
		$g^{(2)}(0)=0.15$ through objective and fiber, respectively.
		\textbf{e}, lifetime measurements through the objective (upper part) and 
		through the fiber end (lower part). Fits to the data (red line) yield 
		lifetimes of \unit{19.7}{\nano\second} and \unit{19.4}{\nano\second}, respectively.
		\textbf{f} intensity through the objective (upper part) and through the fiber's end 
		(lower part) measured at the same time. A blinking behavior is visible. Note 
		the near perfect correlation between the blinking in the upper and lower trace.
		}
		\label{fig:chara}
\end{figure}

We first compare the emission properties of the hBN sources into free space using the high 
NA objective lens with the collection through the fiber. Figure~\ref{fig:chara}(a) and~(b) 
show the room-temperature spectra of the light collected through the objective lens 
and through the fiber, respectively. In both spectra, two peaks are visible (at \unit{573}{\nano\meter} and \unit{666}{\nano\meter}), most likely 
corresponding to two individual single emitters. In the following, the emitter 
at \unit{666}{\nano\meter} will be investigated in detail. Unless otherwise stated, a bandpass 
filter between \unit{650}{\nano\meter} and \unit{700}{\nano\meter} is used. The full 
width at half maximum (FWHM) of the peak is approximately \unit{2}{\nano\meter}. 
Notably, in both cases, namely through the fiber and via the high NA objective, 
the spectra are very similar, and the sharp lines of the emitter is clearly visible. The minor differences can be 
explained by different coupling efficiencies to the fiber and the objective lens and by fluorescence coming from 
pump light entering the silica fiber. 

To prove that the emitter is indeed a single photon source, the second order autocorrelation function, 
$g^{(2)}(t)$, is recorded. Figure~\ref{fig:chara}(c) and~(d) show these antibunching measurements of the light collected through the 
objective lens and through the fiber's end, respectively. The 
measurements were carried out using the \unit{10}{\nano\meter} bandpass filter in order 
to only collect light from the sharp line at \unit{666}{\nano\meter}. Both curves show 
a pronounced dip at zero time delay (t=0), confirming the single photon nature of the light. This clearly proves that  
the coupling of a defect in a two-dimensional material to the fiber and extraction of 
its photons is successful.

In Figure~\ref{fig:chara}(e) lifetime measurements for the free space collection and collection via the fiber are shown. The lifetimes 
through the objective and the fiber are \unit{19.7}{\nano\second} and \unit{19.4}{\nano\second}, respectively.
As expected, these lifetimes are nearly identical, as it is the same emitter. The small differences 
can again be explained by different contributions of background light. To avoid most of these
contributions, the fit was carried out to the data starting from \unit{5}{\nano\second} after the 
excitation laser pulse. 

Finally, we analyze the temporal stability of this emitter by recording the intensity count rate 
as a function of time. The corresponding measurements are shown in Figure~\ref{fig:chara}(f). 
As expected, the emission is modulated in a very similar manner when collected via both outputs, 
namely air objective (upper part)  and fiber (lower part). 
The observed sharp steps in both cases are an indication that this modulation is not caused by the fiber 
vibrating and drifting through the laser focus, where the modulation would be expected to be continuous.
This indicates a blinking behavior of the emitter, that was also be found for some emitters on the substrate.

\begin{figure}
	\centering
		\includegraphics{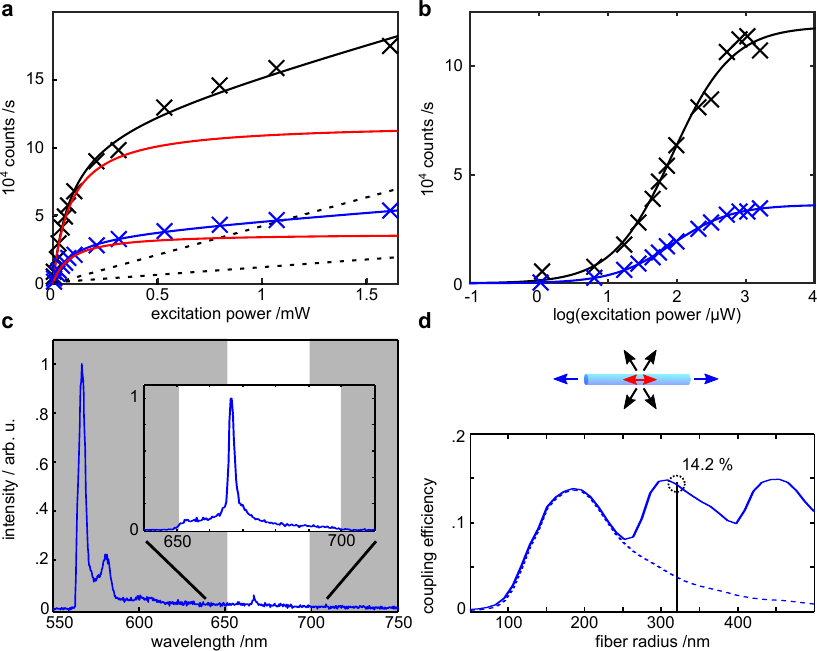}
		\caption{\textbf{Saturation measurements and fiber coupling efficiency.}
		\textbf{a}, Saturation curve of the single defect coupled to the nanofiber measured collecting 
		the light through the objective (black) and through the fiber (blue). 
		The curves follows Equation~\ref{eq:sat} with 
		parameters are $R_{\textrm{inf,obj}}=\unit{118}{\kilo counts\per\second}$, $I_{\textrm{s,obj}}=\unit{86}{\micro\watt}$, and 
		$BG_{\textrm{obj}}=\unit{42}{counts\per\micro\watt}$ for the black curve and 
		$R_{\textrm{inf,fib}}=\unit{36}{\kilo counts\per\second}$, $I_{\textrm{s,fib}}=\unit{86}{\micro\watt}$, and 
		$BG_{\textrm{fib}}=\unit{11}{counts\per\micro\watt}$ for the blue curve, respectively.
		Integration time for each data point was \unit{3}{\second}.
		The red curves are the contribution of the defect whereas the dashed lines are the background contribution.
		\textbf{b}, Representation of the data in \textbf{a,b} on a linear/logarithmic plot with background correction applied.
		\textbf{c}, Spectrum collected when detecting and pumping through the fiber. Even though the 
		pump light propagated through the fiber and excited background fluorescence, the defect at \unit{666}{\nano\meter} is visible. The inset shows 
		the signal after optimization of coupling of the pump light. Shaded areas are filtered out in the inset.
		\textbf{d}, Simulation on the nanofiber diameter and coupling efficiency. The upper part shown a sketch of the 
		geometry investigated while the lower part shows the variation of the coupling into the fiber for 
		a dipole with longitudinal orientation.
		}
		\label{fig:sat}
\end{figure}

To show the high coupling efficiency of the quantum emitter to the fiber, saturation measurements 
were carried out (see Figure~\ref{fig:sat}(a-c)). The saturation curves follow the Equation~\cite{Novotny2012}:
\begin{equation}
R=R_{\inf}\,\frac{\left(\frac{I}{I_{s}}\right)}{\left(\frac{I}{I_{s}}\right)+1} + BG \times I \, ,
\label{eq:sat} 
\end{equation}
with the pump laser intensity $I$, the emission rate at saturation $R_{inf}$, the saturation intensity $I_{s}$, and an additional 
linear background with coefficient $BG$.
From fits to the data saturated emission rates of $R_{\textrm{inf,obj}}=\unit{118}{\kilo counts\per\second}$ 
and $R_{\textrm{inf,fib}}=\unit{36}{\kilo counts\per\second}$ were found for collection through the objective lens 
and one end of the fiber, respectively. For the fit to the confocal data, due to the higher noise, 
the saturation intensity was fixed to $I_{\textrm{s,obj}}=\unit{86}{\micro\watt}$, which is 
the value extracted from the fit to the data through the fiber, that exhibits lower noise levels.

Since the measurements for the count rate at the tapered fiber were only carried out on one end, 
in order to exclude reflections in the open end of the fiber, the reflectivity of the whole fiber (including the reflection from the facet) 
was measured. A value below for the reflectivity below 0.1 was found indicating that this is not the case and the measured count rate 
stems indeed from the photons emitted in one end.

Figure~\ref{fig:sat}(c) shows the emission spectrum, excited with a laser beam that is propagating through the fiber. 
In this configuration, the propagating laser beam can excite background fluorescence inside the fiber. Nevertheless, 
due to a high coupling efficiency for excitation and emitted photons, and the use of a pure silica core tapered fiber, 
the defect's emission can be detected as a bright line at \unit{666}{\nano\meter}. 
Excitation and detection through the fiber will enable the system to serve as an integrated single photon source in quantum networks.

Figure~\ref{fig:sat}d shows the results of numerical simulations of the coupling efficiency for a dipole oriented longitudinal along the fiber 
yielding an efficiency of 14.2~\% for the fiber diameter used in the experiment.

\begin{figure}
	\centering
		\includegraphics{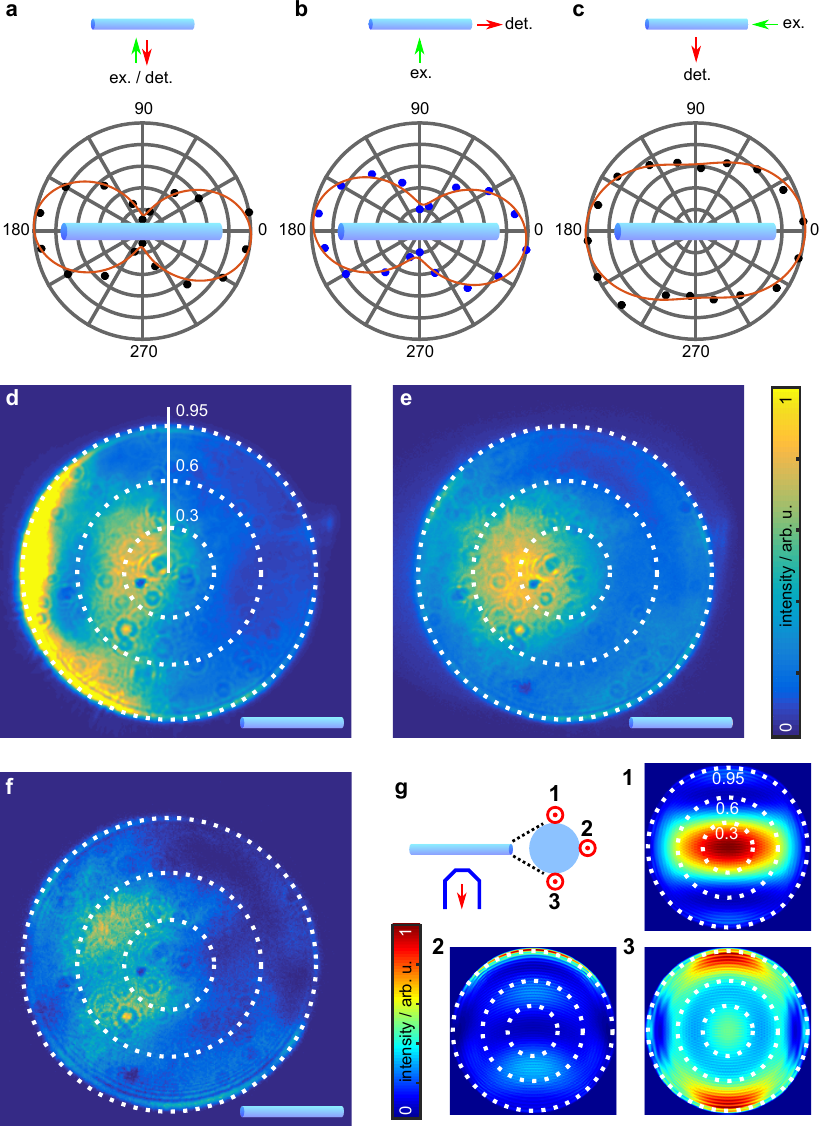}
		\caption{\textbf{Polarisation and back focal plane measurements.}
		\textbf{a}, Intensity of the emission measured through the objective when the excitation polarisation is rotated.
		The fiber axis is horizontal from 0 to 180 degrees. 
		\textbf{b}, Intensity of the emission as measured through the fiber. 
		\textbf{c}, Intensity collected through the objective when excited through the fiber. 
		In contrast to \textbf{a,b}, not the excitation polarisation, but the polarisation of the detection is varied using a polarizer.
		\textbf{d}, Back focal plane (BFP) image when white light from a halogen lamp (\unit{561}{\nano\meter} 
		to \unit{700}{\nano\meter} wavelength) is coupled into the fiber from the right side. Light leaking from the fiber can be seen 
		on the left at high angles while light scattered from the particle attached can be seen in the middle.
		\textbf{e}, BFP image of the fluorescence when pumped through the fiber (\unit{561}{\nano\meter} 
		to \unit{700}{\nano\meter} wavelength). Since the excitation laser gets filtered out, 
		no light leaking from the fiber can be detected. 
		\textbf{f}, BFP image when the \unit{10}{\nano\meter} bandpass filter is introduced. Two lobes 
		are visible separated by a minimum on the fiber axis.
		\textbf{g}, Numerical calculations of the far field pattern and the corresponding collection efficiency for 
		different dipole positions (1: back side, 2: middle position, 3: front side). 
		The best resemblance of the measured pattern was found for a longitudinal 
		dipole at a position at the middle of the fiber.
		The fiber is oriented horizontally in all panels.
		}
		\label{fig:pol}
\end{figure}

To get more insight into the coupling of hBN defect to the nanofiber, we performed measurements of 
the polarisation dependence of excitation and emission. The corresponding measurements are shown in 
Figure~\ref{fig:pol}(a-c). In Figure~\ref{fig:pol}(a,b), the polarisation of the excitation laser is rotated
and the emission collected through the objective and through the fiber, respectively. 
The observed pattern suggests that the transition dipole excited by the laser lies in parallel with the 
nanofiber. 
Figure~\ref{fig:pol}(c) shows the polarisation of the photons collected through the objective when the pump laser 
is coupled into the fiber (i.e. exciting the defect through the fiber). Here, the polarisation has 
a reduced visibility, which could be an indication that the line at \unit{666}{\nano\meter} is 
excited indirectly~\cite{Jungwirth2016}.

By coupling single photon emitters to photonic structures, such as the nanofiber used here, not only new decay 
channels are provided, but also the overall emission pattern is altered. This can be investigated by imaging 
the back focal plane (BFP) of the microscope objective. In the BFP, the angle of the emission to the optical axis 
is directly visible. Figure~\ref{fig:pol}(d-f) shows BFP images of the nanofiber coupled with he hBN flake 
in three different configurations.
First, in Figure~\ref{fig:pol}(d), in order to check the general 
scattering properties, broadband light from a halogen 
lamp is coupled into the fiber and the scattered light observed. Light leaking from the fiber is visible under 
high angles at the left side of the image. In addition, scattering from the particle hosting the hBN flake can be seen in the middle. Second, 
in Figure~\ref{fig:pol}(e), the BFP image is shown when exciting through the fiber. Due to contributions of other sources of light at other wavelengths  
a near isotropic emission is found. 
Finally, to only investigate the light from the single defect at \unit{666}{\nano\meter} wavelength,
in Figure~\ref{fig:pol}(f) the fluorescence when using the \unit{10}{\nano\meter} bandpass 
is shown. Two lobes are visible, separated by a minimum along the fiber axis. This situation 
can be compared to the numerical calculation shown in Figure~\ref{fig:pol}(g) (see Methods). 

In these calculations, the back focal plane intensity pattern is investigated for a dipole 
in parallel orientation  
at positions on top of the fiber, at the middle position, and at the bottom position (labeled 1-3 respectively).
The data shown is is normalized to each panel's maximum and the fraction of the total emitted power 
falling in the numerical aperture of 0.95 used in the experiment is 0.22, 0.22, and 0.51, respectively. 
This shows that the position of the dipole has a dramatic effect on the number of photons that can be 
collected with an objective lens in free space that has to be taken into account when calculation the 
coupling efficiency to the nanofiber. The assumption of isotropic emission of the emitter, 
while often made, is a relatively coarse approximation, as we will investigate later
in this paper.
In our case, the pattern when the dipole sits at the middle of the fiber is closest to the measured result in 
Figure~\ref{fig:pol}(f). Hence, the calculated free space collection efficiency is $\eta_{\textrm{obj,calc}}=0.22$. 

The values found for the collection efficiency differ by more than a factor of two for different dipole positions.
It is therefore illustrative to compare them with the result when assuming isotropic emission of the dipole.
In this case,  
the 0.95 NA collection lens collects $\eta_{\textrm{obj,geom}} \approx \unit{34}{\%}$ of the full solid angle. This would 
-- using the value of  $\eta_{\textrm{fib,calc}}=14.2~\%$ for the coupling efficiency to the fiber
from the calculations shown in Figure~\ref{fig:sat}(d) -- yield a value for the microscope collection efficiency of 
$\eta_{\textrm{obj,iso}}=(1-\eta_{\textrm{fib,calc}}) \times 34~\% = 0.29$.

In order to get an experimental value for the coupling efficiency in the fiber, the transmission of the 
optics has to be known (see Methods). Experimentally, we found a relative transmission for the 
two different paths of $t_{\textrm{rel}}=\frac{t_{\textrm{obj}}}{t_{\textrm{fib}}}=0.72$.
The coupling efficiency to the fiber $\eta_{\textrm{fib}}$ is then given by:
\begin{equation}
\eta_{\textrm{fib}} = \frac{R_{\textrm{inf,fib}} \times 2}{R_{\textrm{inf,obj}}} \times t_{\textrm{rel}} \times \eta_{\textrm{obj,calc}} = 10~\% \, ,
\end{equation}
where the factor of 2 accounts for the fact that light is emitted to both ends of the fiber.

This shows, that the fiber coupled system 
is able to achieve a similar collection efficiency as a high NA objective lens, while having the advantage of being directly 
fiber coupled and offers alignment free collection. With the use of nanofiber Bragg cavities, this value can 
even surpass high NA objective lenses and efficiencies over 80~\% are expected~\cite{Takashima2016}.

The value of the coupling efficiency of 10~\% found here is consistent with the findings of other results using other emitters such as 
quantum dots~\cite{Fujiwara2011,Yalla2012} or defect centers~\cite{Schroeder2012,Liebermeister2014}. 
While those measurements were performed using a single mode nanofiber, here a fibre in the multimode regime was used.  
In this regime the total coupling efficiency can be even higher than in the single mode 
regime~\cite{Kumar2015,Shi2016}, what can be seen in Figure~\ref{fig:sat}(d). For the peak for small fiber diameters 
(single mode regime) is smaller than the other two peaks for a larger radius.

In conclusion, we realized an integrated photonic system where a quantum emitter in a 2D material, 
layered hBN is coupled to a tapered optical fiber. We achieved highly efficient coupling to the fiber with an efficiency of 10~\% 
and were able to prove the quantum nature of the light emitted. 
BFP imaging of the emission dipole as well as polarization measurements confirm the efficient 
coupling of the quantum emitter to the tapered fiber and show that the dipolar emission 
pattern of the emitter has to be taken into account when measuring the coupling efficiency.
The highly efficient coupling observed even allows for excitation of the emitter and collection 
through the same fiber, establishing a fully fiber integrated system.
Our results pave the way for promising applications in integrated quantum 
photonics using single emitters in hBN.

\section*{Acknowledgment} 
We thank Kali Nayak for help with the low fluorescent nanofibers, Syun Suezawa for help with 
Figure~1 and Hironaga Maruya and Atsushi Fukuda for support.
The work was further supported by MEXT/JSPS KAKENHI Grant Number 26220712, 21102007, JST CREST project, and 
Special Coordination Funds for Promoting Science and Technology
AWS thanks funding by the Japanese Society for the Promotion of Science through a fellowship 
for overseas researchers. 
Financial support from the Australian Research Council (IH150100028, DE130100592) and the Asian Office of Aerospace 
Research and Development grant FA2386-15-1-4044 are gratefully acknowledged.

\section*{Methods} 
\textbf{hBN samples}\\
A native oxide Si (100) substrate was cleaned thoroughly with acetone, isopropanol, and ethanol before
drop-casting \unit{100}{\micro\liter} of ethanol solution containing pristine hBN flakes (Graphene Supermarket) of
approximately \unit{200}{\nano\meter} in diameter and  \unit{5 - 50}{\nano\meter} 
of height onto silicon substrates. The completely dried
sample was then loaded into a fused-quartz tube in a tube furnace (Lindberg Blue). The tube was
evacuated to low vacuum ( \unit{10^{-3}}{Torr}) by using a rough pump then purged for 
\unit{30}{\minute} under \unit{50}{sccm} of Ar
with pressure regulated at \unit{1}{Torr}. The substrate was then annealed at
\unit{850}{\Celsius} for \unit{30}{\minute} under \unit{1}{Torr} of
argon. This thermal treatment step is employed to increase the optically active defect density~\cite{Tran2016a}.

\textbf{Tapered fibers}\\
The tapered fiber is fabricated by heating an optical fiber (0.25Z-U, Sumitomo Electric) with a ceramic heater
to a temperature of about \unit{1350}{\degreecelsius}.
The fiber was stretched for \unit{200}{\second} with speeds from \unit{0.2-0.3}{\milli\meter\per\second}
resulting in an overall distance of \unit{505}{\milli\meter}.
With this, a waist diameter of about \unit{640}{\nano\meter} was achieved.
The transmission of the fiber stayed (apart from mode beating expected in multimode fibers) 
close to unity during this process. 

\textbf{Setup}\\
The experimental setup used (shown in Figure~\ref{fig:setup}) consists of a home build confocal microscope and a 
manipulator. The manipulator consists of a sharp tungsten tip (TP-0002, Micro Support) and a three axis piezo stage (TRITOR 100, Piezosystem Jena). 
The microscope objective used (MPlanApo N 100x/0.95, Olympus) has a numerical aperture of 0.95 and is mounted on a one axis piezo stage 
(TRITOR 100 SG, Piezosystem Jena) to adjust the focus. Scanning is achieved by a scanning mirror unit (GVS002, Thorlabs) and the 
excitation lasers used  are a \unit{532}{\nano\meter} wavelength continuous wave laser (LBX-532, Oxxius) and a picosecond 
laser at the same wavelength (Katana, Onefive) for pulsed measurements. Detection is 
done via two avalanche photodiodes (SPCMM-AQRH-14-FC, Perkin Elmer) behind a fiber beam splitter connected to a counting module 
(Time Harp 200, Picoquant) or by a Peltier cooled camera (DU420-OE, Andor) behind a monochromator (Oriel MS257). For 
back focal plane imaging, the back focal plane is imaged on a Peltier cooled camera (PIXIS 1024, Princeton Instruments). 
Transmissions of the microscope and fiber setup was measured to be 0.56 (without \unit{10}{\nano\meter} bandpass filter,
fiber in-coupling, and microscope objective) and 0.7 for the setup used for filtering the light from the fiber 
(not including fiber in- and out-coupling).
 The transmission of the bandpass filter cancels out as it 
is used in both setups and the difference in fiber coupling is estimated to be {90}~{\%}. From the 
manufacturer, the transmission of the transmission of the microscope objective is given as {90}~{\%} 
by the manufacturer. The relative transmission is hence $t_{rel}=\frac{t_{\textrm{c}}}{t_{\textrm{f}}}=0.72$.

\textbf{FDTD simulations} \\
In order to analyze the far-field pattern and the corresponding collection efficiency, 
three dimensional finite-difference time-domain (FDTD) simulations were performed on a commercial package 
(FDTD Solutions, Lumerical). The calculation region (length $\times$ width $\times$ height) were set to \unit{40}{\micro\meter} 
$\times$ \unit{40}{\micro\meter} $\times$ \unit{4}{\micro\meter}. An automatic nonuniform mesh and the material properties of 
SiO2 provided by the software were used in the calculation. A single dipole source was slightly outside 
(\unit{0.5}{\nano\meter}) the surface of the fiber. Perfectly matched layers (PML) were employed as absorbing boundary 
conditions. The far-field pattern was calculated from the top side of the calculation region. 

The coupling efficiency to the nanofiber is calculated by comparing the total emission a dipole with the 
power coupled into the fiber measured with a $\unit{4}{\micro\meter} \times \unit{4}{\micro\meter}$ sized 
monitor at the end of the nanofiber.

\bibliography{bib4}

\providecommand{\latin}[1]{#1}
\makeatletter
\providecommand{\doi}
  {\begingroup\let\do\@makeother\dospecials
  \catcode`\{=1 \catcode`\}=2\doi@aux}
\providecommand{\doi@aux}[1]{\endgroup\texttt{#1}}
\makeatother
\providecommand*\mcitethebibliography{\thebibliography}
\csname @ifundefined\endcsname{endmcitethebibliography}
  {\let\endmcitethebibliography\endthebibliography}{}
\begin{mcitethebibliography}{33}
\providecommand*\natexlab[1]{#1}
\providecommand*\mciteSetBstSublistMode[1]{}
\providecommand*\mciteSetBstMaxWidthForm[2]{}
\providecommand*\mciteBstWouldAddEndPuncttrue
  {\def\EndOfBibitem{\unskip.}}
\providecommand*\mciteBstWouldAddEndPunctfalse
  {\let\EndOfBibitem\relax}
\providecommand*\mciteSetBstMidEndSepPunct[3]{}
\providecommand*\mciteSetBstSublistLabelBeginEnd[3]{}
\providecommand*\EndOfBibitem{}
\mciteSetBstSublistMode{f}
\mciteSetBstMaxWidthForm{subitem}{(\alph{mcitesubitemcount})}
\mciteSetBstSublistLabelBeginEnd
  {\mcitemaxwidthsubitemform\space}
  {\relax}
  {\relax}

\bibitem[Mak and Shan(2016)Mak, and Shan]{Mak2016}
Mak,~K.~F.; Shan,~J. Photonics and optoelectronics of 2D semiconductor
  transition metal dichalcogenides. \emph{Nature Photonics} \textbf{2016},
  \emph{10}, 216--226\relax
\mciteBstWouldAddEndPuncttrue
\mciteSetBstMidEndSepPunct{\mcitedefaultmidpunct}
{\mcitedefaultendpunct}{\mcitedefaultseppunct}\relax
\EndOfBibitem
\bibitem[Novoselov \latin{et~al.}(2016)Novoselov, Mishchenko, Carvalho, and
  Neto]{Novoselov2016}
Novoselov,~K.; Mishchenko,~A.; Carvalho,~A.; Neto,~A.~C. 2D materials and van
  der Waals heterostructures. \emph{Science} \textbf{2016}, \emph{353},
  aac9439\relax
\mciteBstWouldAddEndPuncttrue
\mciteSetBstMidEndSepPunct{\mcitedefaultmidpunct}
{\mcitedefaultendpunct}{\mcitedefaultseppunct}\relax
\EndOfBibitem
\bibitem[Wu \latin{et~al.}(2015)Wu, Buckley, Schaibley, Feng, Yan, Mandrus,
  Hatami, Yao, Vuckovic, Majumdar, and et~al.]{Wu2015}
Wu,~S.; Buckley,~S.; Schaibley,~J.~R.; Feng,~L.; Yan,~J.; Mandrus,~D.~G.;
  Hatami,~F.; Yao,~W.; Vuckovic,~J.; Majumdar,~A.; et~al., Monolayer
  semiconductor nanocavity lasers with ultralow thresholds. \emph{Nature}
  \textbf{2015}, \emph{520}, 69--72\relax
\mciteBstWouldAddEndPuncttrue
\mciteSetBstMidEndSepPunct{\mcitedefaultmidpunct}
{\mcitedefaultendpunct}{\mcitedefaultseppunct}\relax
\EndOfBibitem
\bibitem[Caldwell \latin{et~al.}(2014)Caldwell, Kretinin, Chen, Giannini,
  Fogler, Francescato, Ellis, Tischler, Woods, Giles, and et~al.]{Caldwell2014}
Caldwell,~J.~D.; Kretinin,~A.~V.; Chen,~Y.; Giannini,~V.; Fogler,~M.~M.;
  Francescato,~Y.; Ellis,~C.~T.; Tischler,~J.~G.; Woods,~C.~R.; Giles,~A.~J.;
  et~al., Sub-diffractional volume-confined polaritons in the natural
  hyperbolic material hexagonal boron nitride. \emph{Nature communications}
  \textbf{2014}, \emph{5}, 5221--5221\relax
\mciteBstWouldAddEndPuncttrue
\mciteSetBstMidEndSepPunct{\mcitedefaultmidpunct}
{\mcitedefaultendpunct}{\mcitedefaultseppunct}\relax
\EndOfBibitem
\bibitem[Dai \latin{et~al.}(2014)Dai, Fei, Ma, Rodin, Wagner, McLeod, Liu,
  Gannett, Regan, Watanabe, and et~al.]{Dai2014}
Dai,~S.; Fei,~Z.; Ma,~Q.; Rodin,~A.; Wagner,~M.; McLeod,~A.; Liu,~M.;
  Gannett,~W.; Regan,~W.; Watanabe,~K.; et~al., Tunable phonon polaritons in
  atomically thin van der Waals crystals of boron nitride. \emph{Science}
  \textbf{2014}, \emph{343}, 1125--1129\relax
\mciteBstWouldAddEndPuncttrue
\mciteSetBstMidEndSepPunct{\mcitedefaultmidpunct}
{\mcitedefaultendpunct}{\mcitedefaultseppunct}\relax
\EndOfBibitem
\bibitem[Yang \latin{et~al.}(2016)Yang, Shang, Wang, Shen, Cao, Peimyoo, Zou,
  Chen, Wang, Cong, and et~al.]{Yang2016}
Yang,~W.; Shang,~J.; Wang,~J.; Shen,~X.; Cao,~B.; Peimyoo,~N.; Zou,~C.;
  Chen,~Y.; Wang,~Y.; Cong,~C.; et~al., Electrically tunable valley-light
  emitting diode (vLED) based on CVD-grown monolayer WS2. \emph{Nano letters}
  \textbf{2016}, \emph{16}, 1560--1567\relax
\mciteBstWouldAddEndPuncttrue
\mciteSetBstMidEndSepPunct{\mcitedefaultmidpunct}
{\mcitedefaultendpunct}{\mcitedefaultseppunct}\relax
\EndOfBibitem
\bibitem[Sie \latin{et~al.}(2015)Sie, McIver, Lee, Fu, Kong, and
  Gedik]{Sie2015}
Sie,~E.~J.; McIver,~J.~W.; Lee,~Y.-H.; Fu,~L.; Kong,~J.; Gedik,~N.
  Valley-selective optical Stark effect in monolayer WS2. \emph{Nature
  materials} \textbf{2015}, \emph{14}, 290--294\relax
\mciteBstWouldAddEndPuncttrue
\mciteSetBstMidEndSepPunct{\mcitedefaultmidpunct}
{\mcitedefaultendpunct}{\mcitedefaultseppunct}\relax
\EndOfBibitem
\bibitem[Clark \latin{et~al.}(2016)Clark, Schaibley, Ross, Taniguchi, Watanabe,
  Hendrickson, Mou, Yao, and Xu]{Clark2016}
Clark,~G.; Schaibley,~J.~R.; Ross,~J.~S.; Taniguchi,~T.; Watanabe,~K.;
  Hendrickson,~J.~R.; Mou,~S.; Yao,~W.; Xu,~X. Single Defect Light Emitting
  Diode in a van der Waals Heterostructure. \emph{Nano letters} \textbf{2016},
  \emph{16}, 3944--3948\relax
\mciteBstWouldAddEndPuncttrue
\mciteSetBstMidEndSepPunct{\mcitedefaultmidpunct}
{\mcitedefaultendpunct}{\mcitedefaultseppunct}\relax
\EndOfBibitem
\bibitem[Withers \latin{et~al.}(2015)Withers, Del Pozo-Zamudio, Mishchenko,
  Rooney, Gholinia, Watanabe, Taniguchi, Haigh, Geim, Tartakovskii, and
  et~al.]{Withers2015}
Withers,~F.; Del Pozo-Zamudio,~O.; Mishchenko,~A.; Rooney,~A.; Gholinia,~A.;
  Watanabe,~K.; Taniguchi,~T.; Haigh,~S.; Geim,~A.; Tartakovskii,~A.; et~al.,
  Light-emitting diodes by band-structure engineering in van der Waals
  heterostructures. \emph{Nature materials} \textbf{2015}, \emph{14},
  301--306\relax
\mciteBstWouldAddEndPuncttrue
\mciteSetBstMidEndSepPunct{\mcitedefaultmidpunct}
{\mcitedefaultendpunct}{\mcitedefaultseppunct}\relax
\EndOfBibitem
\bibitem[Tran \latin{et~al.}(2016)Tran, ElBadawi, Totonjian, Lobo, Grosso,
  Moon, Englund, Ford, Aharonovich, and Toth]{Tran2016}
Tran,~T.~T.; ElBadawi,~C.; Totonjian,~D.; Lobo,~C.~J.; Grosso,~G.; Moon,~H.;
  Englund,~D.~R.; Ford,~M.~J.; Aharonovich,~I.; Toth,~M. Robust multicolor
  single photon emission from point defects in hexagonal boron nitride.
  \emph{ACS Nano} \textbf{2016}, \emph{10}, 7331--7338\relax
\mciteBstWouldAddEndPuncttrue
\mciteSetBstMidEndSepPunct{\mcitedefaultmidpunct}
{\mcitedefaultendpunct}{\mcitedefaultseppunct}\relax
\EndOfBibitem
\bibitem[Bourrellier \latin{et~al.}(2016)Bourrellier, Meuret, Tararan, Stephan,
  Kociak, G.~Tizei, and Zobelli]{Bourrellier2016}
Bourrellier,~R.; Meuret,~S.; Tararan,~A.; Stephan,~O.; Kociak,~M.;
  G.~Tizei,~L.~H.; Zobelli,~A. Bright UV single photon emission at point
  defects in h-BN. \emph{Nano letters} \textbf{2016}, \emph{16},
  1317--4321\relax
\mciteBstWouldAddEndPuncttrue
\mciteSetBstMidEndSepPunct{\mcitedefaultmidpunct}
{\mcitedefaultendpunct}{\mcitedefaultseppunct}\relax
\EndOfBibitem
\bibitem[Jungwirth \latin{et~al.}(2016)Jungwirth, Calderon, Ji, Spencer, Flatt,
  and Fuchs]{Jungwirth2016}
Jungwirth,~N.~R.; Calderon,~B.; Ji,~Y.; Spencer,~M.~G.; Flatt,~M.~E.;
  Fuchs,~G.~D. Temperature Dependence of Wavelength Selectable Zero-Phonon
  Emission from Single Defects in Hexagonal Boron Nitride. \emph{Nano Lett.}
  \textbf{2016}, \emph{16}, 6052--6057\relax
\mciteBstWouldAddEndPuncttrue
\mciteSetBstMidEndSepPunct{\mcitedefaultmidpunct}
{\mcitedefaultendpunct}{\mcitedefaultseppunct}\relax
\EndOfBibitem
\bibitem[Chejanovsky \latin{et~al.}(2016)Chejanovsky, Rezai, Paolucci, Kim,
  Rendler, Rouabeh, Fa?varo~de Oliveira, Herlinger, Denisenko, Yang, and
  et~al.]{Chejanovsky2016}
Chejanovsky,~N.; Rezai,~M.; Paolucci,~F.; Kim,~Y.; Rendler,~T.; Rouabeh,~W.;
  Fa?varo~de Oliveira,~F.; Herlinger,~P.; Denisenko,~A.; Yang,~S.; et~al.,
  Structural Attributes and Photodynamics of Visible Spectrum Quantum Emitters
  in Hexagonal Boron Nitride. \emph{Nano Letters} \textbf{2016}, \emph{16},
  7037--7045\relax
\mciteBstWouldAddEndPuncttrue
\mciteSetBstMidEndSepPunct{\mcitedefaultmidpunct}
{\mcitedefaultendpunct}{\mcitedefaultseppunct}\relax
\EndOfBibitem
\bibitem[Schell \latin{et~al.}(2016)Schell, Tran, Takashima, Takeuchi, and
  Aharonovich]{Schell2016}
Schell,~A.~W.; Tran,~T.~T.; Takashima,~H.; Takeuchi,~S.; Aharonovich,~I.
  Non-linear excitation of quantum emitters in hexagonal boron nitride
  multiplayers. \emph{APL Photonics} \textbf{2016}, \emph{1},
  091302--091302\relax
\mciteBstWouldAddEndPuncttrue
\mciteSetBstMidEndSepPunct{\mcitedefaultmidpunct}
{\mcitedefaultendpunct}{\mcitedefaultseppunct}\relax
\EndOfBibitem
\bibitem[Helmchen and Denk(2005)Helmchen, and Denk]{Helmchen2005}
Helmchen,~F.; Denk,~W. Deep tissue two-photon microscopy. \emph{Nature methods}
  \textbf{2005}, \emph{2}, 932--940\relax
\mciteBstWouldAddEndPuncttrue
\mciteSetBstMidEndSepPunct{\mcitedefaultmidpunct}
{\mcitedefaultendpunct}{\mcitedefaultseppunct}\relax
\EndOfBibitem
\bibitem[Gaponenko(2010)]{Gaponenko2010}
Gaponenko,~S.~V. \emph{Introduction to nanophotonics}; Cambridge University
  Press, 2010; pp~--\relax
\mciteBstWouldAddEndPuncttrue
\mciteSetBstMidEndSepPunct{\mcitedefaultmidpunct}
{\mcitedefaultendpunct}{\mcitedefaultseppunct}\relax
\EndOfBibitem
\bibitem[Weber \latin{et~al.}(2010)Weber, Koehl, Varley, Janotti, Buckley,
  Van~de Walle, and Awschalom]{Weber2010}
Weber,~J.; Koehl,~W.; Varley,~J.; Janotti,~A.; Buckley,~B.; Van~de Walle,~C.;
  Awschalom,~D.~D. Quantum computing with defects. \emph{Proceedings of the
  National Academy of Sciences} \textbf{2010}, \emph{107}, 8513--8518\relax
\mciteBstWouldAddEndPuncttrue
\mciteSetBstMidEndSepPunct{\mcitedefaultmidpunct}
{\mcitedefaultendpunct}{\mcitedefaultseppunct}\relax
\EndOfBibitem
\bibitem[Aharonovich \latin{et~al.}(2016)Aharonovich, Englund, and
  Toth]{Aharonovich2016}
Aharonovich,~I.; Englund,~D.; Toth,~M. Solid-state single-photon emitters.
  \emph{Nature Photonics} \textbf{2016}, \emph{10}, 631--641\relax
\mciteBstWouldAddEndPuncttrue
\mciteSetBstMidEndSepPunct{\mcitedefaultmidpunct}
{\mcitedefaultendpunct}{\mcitedefaultseppunct}\relax
\EndOfBibitem
\bibitem[Mansfield and Kino(1990)Mansfield, and Kino]{Mansfield1990}
Mansfield,~S.~M.; Kino,~G.~S. Solid immersion microscope. \emph{Applied Physics
  Letters} \textbf{1990}, \emph{57}, 2615--2616\relax
\mciteBstWouldAddEndPuncttrue
\mciteSetBstMidEndSepPunct{\mcitedefaultmidpunct}
{\mcitedefaultendpunct}{\mcitedefaultseppunct}\relax
\EndOfBibitem
\bibitem[Lee \latin{et~al.}(2011)Lee, Chen, Eghlidi, Kukura, Lettow, Renn,
  Sandoghdar, and Götzinger]{Lee2011}
Lee,~K.~G.; Chen,~X.~W.; Eghlidi,~H.; Kukura,~P.; Lettow,~R.; Renn,~A.;
  Sandoghdar,~V.; Götzinger,~S. A planar dielectric antenna for directional
  single-photon emission and near-unity collection efficiency. \emph{Nature
  Photonics} \textbf{2011}, \emph{5}, 166--169\relax
\mciteBstWouldAddEndPuncttrue
\mciteSetBstMidEndSepPunct{\mcitedefaultmidpunct}
{\mcitedefaultendpunct}{\mcitedefaultseppunct}\relax
\EndOfBibitem
\bibitem[Yalla \latin{et~al.}(2012)Yalla, Le~Kien, Morinaga, and
  Hakuta]{Yalla2012}
Yalla,~R.; Le~Kien,~F.; Morinaga,~M.; Hakuta,~K. Efficient channeling of
  fluorescence photons from single quantum dots into guided modes of optical
  nanofiber. \emph{Physical review letters} \textbf{2012}, \emph{109},
  063602--063602\relax
\mciteBstWouldAddEndPuncttrue
\mciteSetBstMidEndSepPunct{\mcitedefaultmidpunct}
{\mcitedefaultendpunct}{\mcitedefaultseppunct}\relax
\EndOfBibitem
\bibitem[Fujiwara \latin{et~al.}(2011)Fujiwara, Toubaru, Noda, Zhao, and
  Takeuchi]{Fujiwara2011}
Fujiwara,~M.; Toubaru,~K.; Noda,~T.; Zhao,~H.-Q.; Takeuchi,~S. Highly Efficient
  Coupling of Photons from Nanoemitters into Single-Mode Optical Fibers.
  \emph{Nano Letters} \textbf{2011}, \emph{11}, 4362--4365\relax
\mciteBstWouldAddEndPuncttrue
\mciteSetBstMidEndSepPunct{\mcitedefaultmidpunct}
{\mcitedefaultendpunct}{\mcitedefaultseppunct}\relax
\EndOfBibitem
\bibitem[Liebermeister \latin{et~al.}(2014)Liebermeister, Petersen, Munchow,
  Burchardt, Hermelbracht, Tashima, Schell, Benson, Meinhardt, Krueger,
  Stiebeiner, Rauschenbeutel, Weinfurter, and Weber]{Liebermeister2014}
Liebermeister,~L.; Petersen,~F.; Munchow,~A.~v.; Burchardt,~D.;
  Hermelbracht,~J.; Tashima,~T.; Schell,~A.~W.; Benson,~O.; Meinhardt,~T.;
  Krueger,~A.; Stiebeiner,~A.; Rauschenbeutel,~A.; Weinfurter,~H.; Weber,~M.
  Tapered fiber coupling of single photons emitted by a deterministically
  positioned single nitrogen vacancy center. \emph{Applied Physics Letters}
  \textbf{2014}, \emph{104}, 031101--031101\relax
\mciteBstWouldAddEndPuncttrue
\mciteSetBstMidEndSepPunct{\mcitedefaultmidpunct}
{\mcitedefaultendpunct}{\mcitedefaultseppunct}\relax
\EndOfBibitem
\bibitem[Yalla \latin{et~al.}(2014)Yalla, Sadgrove, Nayak, and
  Hakuta]{Yalla2014}
Yalla,~R.; Sadgrove,~M.; Nayak,~K.~P.; Hakuta,~K. Cavity quantum
  electrodynamics on a nanofiber using a composite photonic crystal cavity.
  \emph{Physical review letters} \textbf{2014}, \emph{113},
  143601--143601\relax
\mciteBstWouldAddEndPuncttrue
\mciteSetBstMidEndSepPunct{\mcitedefaultmidpunct}
{\mcitedefaultendpunct}{\mcitedefaultseppunct}\relax
\EndOfBibitem
\bibitem[Schell \latin{et~al.}(2015)Schell, Takashima, Kamioka, Oe, Fujiwara,
  Benson, and Takeuchi]{Schell2015}
Schell,~A.~W.; Takashima,~H.; Kamioka,~S.; Oe,~Y.; Fujiwara,~M.; Benson,~O.;
  Takeuchi,~S. Highly efficient coupling of nanolight emitters to a ultra-wide
  tunable nanofibre cavity. \emph{Scientific reports} \textbf{2015}, \emph{5},
  9619--9619\relax
\mciteBstWouldAddEndPuncttrue
\mciteSetBstMidEndSepPunct{\mcitedefaultmidpunct}
{\mcitedefaultendpunct}{\mcitedefaultseppunct}\relax
\EndOfBibitem
\bibitem[Takashima \latin{et~al.}(2016)Takashima, Fujiwara, Schell, and
  Takeuchi]{Takashima2016}
Takashima,~H.; Fujiwara,~M.; Schell,~A.~W.; Takeuchi,~S. Detailed numerical
  analysis of photon emission from a single light emitter coupled with a
  nanofiber Bragg cavity. \emph{Optics Express} \textbf{2016}, \emph{24},
  15050--15058\relax
\mciteBstWouldAddEndPuncttrue
\mciteSetBstMidEndSepPunct{\mcitedefaultmidpunct}
{\mcitedefaultendpunct}{\mcitedefaultseppunct}\relax
\EndOfBibitem
\bibitem[Fujiwara \latin{et~al.}(2016)Fujiwara, Yoshida, Noda, Takashima,
  Schell, Mizuochi, and Takeuchi]{Fujiwara2016}
Fujiwara,~M.; Yoshida,~K.; Noda,~T.; Takashima,~H.; Schell,~A.~W.;
  Mizuochi,~N.; Takeuchi,~S. Manipulation of single nanodiamonds to ultrathin
  fiber-taper nanofibers and control of NV-spin states toward fiber-integrated
  lambda-systems. \emph{Nanotechnology} \textbf{2016}, \emph{27},
  455202--455202\relax
\mciteBstWouldAddEndPuncttrue
\mciteSetBstMidEndSepPunct{\mcitedefaultmidpunct}
{\mcitedefaultendpunct}{\mcitedefaultseppunct}\relax
\EndOfBibitem
\bibitem[Novotny and Hecht(2012)Novotny, and Hecht]{Novotny2012}
Novotny,~L.; Hecht,~B. \emph{Principles of nano-optics}; Cambridge university
  press, 2012\relax
\mciteBstWouldAddEndPuncttrue
\mciteSetBstMidEndSepPunct{\mcitedefaultmidpunct}
{\mcitedefaultendpunct}{\mcitedefaultseppunct}\relax
\EndOfBibitem
\bibitem[Schroeder \latin{et~al.}(2012)Schroeder, Fujiwara, Noda, Zhao, Benson,
  and Takeuchi]{Schroeder2012}
Schroeder,~T.; Fujiwara,~M.; Noda,~T.; Zhao,~H.-Q.; Benson,~O.; Takeuchi,~S. A
  nanodiamond-tapered fiber system with high single-mode coupling efficiency.
  \emph{Optics express} \textbf{2012}, \emph{20}, 10490--10497\relax
\mciteBstWouldAddEndPuncttrue
\mciteSetBstMidEndSepPunct{\mcitedefaultmidpunct}
{\mcitedefaultendpunct}{\mcitedefaultseppunct}\relax
\EndOfBibitem
\bibitem[Kumar \latin{et~al.}(2015)Kumar, Gokhroo, Deasy, Maimaiti, Frawley,
  Phelan, and Chormaic]{Kumar2015}
Kumar,~R.; Gokhroo,~V.; Deasy,~K.; Maimaiti,~A.; Frawley,~M.~C.; Phelan,~C.;
  Chormaic,~S.~N. Interaction of laser-cooled 87Rb atoms with higher order
  modes of an optical nanofibre. \emph{New Journal of Physics} \textbf{2015},
  \emph{17}, 013026\relax
\mciteBstWouldAddEndPuncttrue
\mciteSetBstMidEndSepPunct{\mcitedefaultmidpunct}
{\mcitedefaultendpunct}{\mcitedefaultseppunct}\relax
\EndOfBibitem
\bibitem[Shi \latin{et~al.}(2016)Shi, Sontheimer, Nikolay, Schell, Fischer,
  Naber, Benson, and Wegener]{Shi2016}
Shi,~Q.; Sontheimer,~B.; Nikolay,~N.; Schell,~A.; Fischer,~J.; Naber,~A.;
  Benson,~O.; Wegener,~M. Wiring up pre-characterized single-photon emitters by
  laser lithography. \emph{Scientific Reports} \textbf{2016}, \emph{6},
  31135--31135\relax
\mciteBstWouldAddEndPuncttrue
\mciteSetBstMidEndSepPunct{\mcitedefaultmidpunct}
{\mcitedefaultendpunct}{\mcitedefaultseppunct}\relax
\EndOfBibitem
\bibitem[Tran \latin{et~al.}(2016)Tran, Bray, Ford, Toth, and
  Aharonovich]{Tran2016a}
Tran,~T.~T.; Bray,~K.; Ford,~M.~J.; Toth,~M.; Aharonovich,~I. Quantum emission
  from hexagonal boron nitride monolayers. \emph{Nature nanotechnology}
  \textbf{2016}, \emph{11}, 37--41\relax
\mciteBstWouldAddEndPuncttrue
\mciteSetBstMidEndSepPunct{\mcitedefaultmidpunct}
{\mcitedefaultendpunct}{\mcitedefaultseppunct}\relax
\EndOfBibitem
\end{mcitethebibliography}

\end{document}